\documentclass[prd,twocolumn,showpacs,preprintnumbers,superscriptaddress]{revtex4}
\usepackage{epsfig}
\usepackage{amsmath}
\begin{document}

 \preprint{RBRC-562}
\preprint{UMDOE40762-351}

\title{Threshold Resummation for Higgs Production in
Effective Field Theory}

\author{Ahmad Idilbi}
\affiliation{Department of Physics, University of Maryland,
College Park, Maryland 20742, USA}
\author{Xiangdong Ji}
\affiliation{Department of Physics, University of Maryland,
College Park, Maryland 20742, USA} \affiliation{Department of
Physics, Peking University, Beijing, 100871, P. R. China}
\affiliation{Institute of Theoretical Physics, Academia Sinica,
Beijing, 100080, P. R. China}
\author{Jian-Ping Ma}
\affiliation{Institute of Theoretical Physics, Academia Sinica,
Beijing, 100080, P. R. China}
 \affiliation{Department of
Physics, Peking University, Beijing, 100871, P. R. China}
\author{Feng Yuan}
\affiliation{RIKEN/BNL Research Center, Building 510A, Brookhaven
National Laboratory, Upton, NY 11973}

\date{\today}
\vspace{0.5in}
\begin{abstract}
We present an effective field theory approach to resum the large
double logarithms originated from soft-gluon radiations at small
final-state hadron invariant masses in Higgs and vector boson
$(\gamma^*, W, Z)$ production at hadron colliders. The approach is
conceptually simple, independent of details of an effective field
theory formulation, and valid to all orders in sub-leading
logarithms. As an example, we show the result of summing the
next-to-next-to-next-to leading logarithms is identical to that of
the standard pQCD factorization method.\\

\end{abstract}
\maketitle

{\bf 1.} {\it Introduction}. In hadron colliders such as Tevatron
and LHC, the rates on Higgs boson and Drell-Yan pair production
demand reliable perturbative quantum chromodynamics (pQCD)
calculations. When the final-state invariant mass of hadrons is
small, a fixed-order pQCD calculation yields large threshold
double logarithms in the coefficient functions
\begin{equation}
               \alpha_s^k \left[\frac{
\ln^{m-1}(1-z)}{(1-z)}\right]_+\ ,   ~~~~(m\le
                2k)
\end{equation}
which must be resummed to all orders in $\alpha_s$, where $1-z$ is
the fraction of center-of-mass energy of the initial partons going
into soft radiations. In moment space, these large logarithms
appear in the form, $\alpha_s^k \ln^m \overline{N}$, where
$\overline{N}=N\exp(\gamma_E)$ with $N$, the order of moment and
$\gamma_E$, the Euler constant. In the past decade, a standard
method based on pQCD factorization has been established to perform
the resummation \cite{Ste87,CatTre89,Forte:2002ni}, and it has
been carried out to next-to-next-to-next to leading logarithms
(N$^3$LL) for the above processes
\cite{drellyan,Catani:2001ic,Anastasiou:2002yz,Catani:2003zt,Moch:2005ba,Moch:2005ky,Laenen:2005uz}.

In this paper, we present an alternative, effective field theory
(EFT) resummation of these large threshold logarithms. The EFT
approach is conceptually simple, and is readily extended to other
processes. It is motivated by the recent development of
soft-collinear effective field theory
\cite{Bauer:2000yr,Bauer:2002nz} and its applications to threshold
resummation \cite{Man03,Idilbi:2005ky}. The major steps described
here are standard in EFT methodology and are similar to those in
Refs. \cite{Man03,Idilbi:2005ky}. However, there are also
significant differences: First, we work in the limit of $Q=M_H$
(Higgs mass) going to infinity first (twist expansion), and
$z\rightarrow 1$ subsequently. The kinematic region $(1-z)\sim
\Lambda_{\rm QCD}/Q$, where there might be subtleties in
formulating an effective theory \cite{Pecjak:2005uh}, is avoided.
Second, our approach uses the full QCD results in the
soft-collinear limit, and thus an actual formulation of an
effective lagrangian is unnecessary. Finally, our resummation
formula is valid to all orders in sub-leading logarithms and fully
equivalent to the standard resummation. As an example, we
demonstrate that the EFT resummation to N$^3$LL agrees completely
with the latest result in the literature using the traditional
approach \cite{Moch:2005ky,Laenen:2005uz}.

Let us consider the standard model Higgs production in
hadron-hadron collisions. An almost identical discussion applies
to Drell-Yan production of $\gamma^*$, $W$ and $Z$ bosons. At high
energy, the Higgs cross section is dominated by gluon-gluon fusion
through the top quark loop. For a large range of Higgs mass, the
top quark can be considered as heavy and can be integrated out
first \cite{Dawson:1990zj}. The effective lagrangian density for
Higgs production is \begin{equation}
    {\cal L} = -\frac{1}{4}C_\phi(M_t,\mu_R) ~\phi
~G^{\mu\nu}G_{\mu\nu}(\mu_R) \ ,
\end{equation}
where $\phi$ is the scalar field, $G^{\mu\nu}$ is the gluon field
strength, $C_\phi$ is the effective coupling
\cite{Chetyrkin:1997iv}, and $\mu_R$ is a renormalization scale.
We focus on the kinematic limit in which the final-state hadron
invariant mass $(1-z)M_H$ is small in the sense that $(1-z)\ll 1$.
The scale $(1-z)M_H$, however, is still perturbative and is in
principle much larger than $\sqrt{\Lambda_{\rm QCD} M_H}$. In the
above kinematic limit, the pQCD process is dominated by soft and
collinear gluon radiations.

Renormaliztion scale $\mu_R$ is arbitrary in principle, but we
will set it to $M_H$ in the following discussion. The process then
has three independent scales: $M_H$, $(1-z)M_H$ and $\mu_F$, last
of which is related to factoring the collinear divergences into
Feynman parton distributions. To calculate the Higgs production
cross section reliably in the threshold region, we shall study the
physics at different scales separately using EFT techniques
\cite{Man03}.

{\bf 2.} {\it Physics at scale $M_H$, and between $M_H$ and
$(1-z)M_H$.} Momentum scale $M_H$ is confined to the
gluon-Higgs-gluon vertex region. To account for their
contribution, we introduce the scalar current $J_{\rm
QCD}(M_H)=G^{\mu\nu}G_{\mu\nu}(M_H)$ and make the operator
expansion, \begin{equation}
     J_{\rm QCD}(M_H)=
C_g(M_H/\mu,\alpha_s(\mu))J_{\rm eff}(\mu) +
     ... \ ,
\end{equation}
where coefficient function $C_g$ contains the perturbative
contribution between momentum scale $M_H$ and $\mu$. $J_{\rm
eff}(\mu)$ contains the soft and collinear contributions below
scale $\mu$. To calculate the coefficient function, we take the
matrix element of the above equation between gluon states---the
left-hand side defines the gluon form factor $F_g(M_H)$.

In dimensional regularization and minimal subtraction scheme, the
gluon form factor can be calculated as a power series in the
renormalized strong coupling,
\begin{equation}
     F_g(M_H) = Z_{G^2}(M_H)\sum_{l=0}^\infty
     \left(a_s(\mu)\right)^l
     \left(\frac{Q^2}{\mu^2}\right)^{-l\epsilon}
F^{(l)}_g \
\end{equation}
where $Z_{G^2}$ is a renormalization constant, $F^{(l)}_g$
contains infrared $1/\epsilon$-poles. Following
\cite{Moch:2005ba}, we introduce $a_s = \alpha_s/4\pi$ as
expansion parameter for the form factor calculated at higher
order. The infrared divergences can be isolated by the following
factorization, \begin{equation}
    F_g(M_H) = C_g\left(M_H/\mu,\alpha_s(\mu)\right)
    S_g\left(M_H/\mu,\alpha_s(\mu), 1/\epsilon\right) \end{equation}
where $S_g$ contains the pole contributions only and can be
regarded as the gluon matrix element of effective current $J_{\rm
eff}(\mu)$. Expanding the coefficient function at $\mu=M_H$ as
$C_g(1,\alpha_s(M_H)) = \sum_i a_s^i(M_H) C^{(i)}_g$, and using
the form factor calculated to the second order in $\alpha_s$
\cite{Harlander:2000mg}, we obtain
\begin{eqnarray}
   C^{(1)}_g &=& 7C_A\zeta_2 \nonumber \\
   C^{(2)}_g &=& C_A^2\left(\frac{5105}{162}
   + \frac{335}{6}\zeta_2 - \frac{143}{9}\zeta_3
      + \frac{125}{10}\zeta^2_2\right) \nonumber \\
      &+& C_An_f\left(-\frac{916}{81} -
\frac{25}{3}\zeta_2
      -\frac{46}{9}\zeta_3\right) \nonumber \\
      &+& C_Fn_f\left(-\frac{67}{6} + 8\zeta_3\right)
\ ,
\end{eqnarray}
where $\zeta_n$ is the Riemann zeta-function, $n_f$ is the number
of massless quark flavor, $C_A=N_c$ and $C_F=(N_c^2-1)/2N_c$ and
$N_c=3$ is the number of color.

The physics between $M_H$ and $M_H(1-z)$ can be accounted for by
integrating over the running scale $\mu$ between them. This
requires an anomalous dimension,
\begin{equation}
              \gamma_{1,g}(\alpha_s) = \mu \frac{d\ln C_g}{d\mu} = -
\mu \frac{d\ln S_g}{d\mu}
              \equiv \sum_i a_s^i \gamma^{(i)}_{1,g} \
.
\end{equation}
A simple way to calculate this is to start with the following
representation of the form factor \cite{Collins:1989bt},
\begin{eqnarray}
\ln F_g(\alpha_s) &=& \frac{1}{2} \int^{M_H^2/\mu^2}_0
\left.\frac{d\xi}{\xi}
   \right(K_g(\alpha_s(\mu),\epsilon)  \\
   &+& \left. G_g(1, \alpha_s(\xi\mu,\epsilon),
\epsilon) + \int^1_\xi
     \frac{d\lambda}{\lambda} A_g(\alpha_s(\lambda
\mu,\epsilon)\right)
     \nonumber
\label{ano1}
\end{eqnarray}
where $A_g$ is the cusp anomalous dimension of Wilson lines in
adjoint representation \cite{Korchemsky:1987wg}, $A_g=\sum_i
a_s^iA_g^{(i)}$, and has been calculated to three-loops recently
\cite{Vogt:2004mw}. The soft function $K_g$ contains only the
infrared pole terms and can be constructed from $A_g$ through
$\mu^2dK_g/d\mu^2=-A_g$. The perturbative function $G_g = \sum
a_s^i G^{(i)}_g$ contains both finite terms and those vanishing
when $\epsilon\rightarrow 0$. According to Ref.
\cite{Moch:2005tm}, $G^{(i)}_g$ has a simple representation
\begin{equation}
     G^{(i)}_g = 2B^{(i)}_{2,g} - 2i\beta_{i-1} +
f_g^{(i)} + \Delta
     G^{(i)}_g \ ,
\label{ano2}
\end{equation}
for $i=1,~2,~3$, where $\Delta G^{(i)}_g$ can be constructed from
those vanishing terms in lower-order $G^{(i)}_g$.
$B_{2,g}=\sum_ia_s^iB_{2,g}^{(i)}$ is the coefficient of
$\delta(1-x)$ term in the gluon splitting function, with
$B^{(1)}_{2,g} = 11C_A/3-2n_f/3$, etc \cite{Vogt:2004mw}. The QCD
$\beta$-function is defined as $\beta(a_s) =-d \ln
\alpha_s/d\ln\mu^2 = \beta_0a_s + \beta_1a_s^2 + ..$, with
$\beta_0 = 11C_A/3-2n_f/3$. The functions $f_g^{(i)}$ are
universal in the sense that the corresponding quark expressions
are obtained by replacing the overall factor of $C_A$ by $C_F$.

The pole terms in Eq.~(8), the $S_g$ terms, will be used to
calculate the anomalous dimenstion by employing the last two
relations of Eq.~(7). After taking into account the
renormalization of the strong coupling constant $\alpha_s(\mu)$,
we get the anomalous dimension of the gluon current as
\begin{equation}
     \gamma^{(i)}_{1,g} = A^{(i)}_g\ln \left(
M^2_H/\mu^2\right) +
     B^{(i)}_{1,g}+2i\beta_{i-1} \ ,
\end{equation}
where
\begin{equation}
   B^{(i)}_{1,g} =-2 B^{(i)}_{2,g}- f_g^{(i)} \ . \end{equation} Since
$A^{(i)}$, $B^{(i)}_{2,g}$, and $f_g^{(i)}$ are known to three
loops \cite{Ravindran:2004mb,Moch:2005tm}, the anomalous dimension
is now known to the same order.

Using the evolution equation, one can summarize the physics
between scales $M_H$ and $\mu_I\sim (1-z)M_H$ in $C_g(M_H/\mu,
\alpha_s(\mu))$ with a coefficient $C_g(1, \alpha_s(M_H))$ and
running
\begin{eqnarray}
 C_g\left(\frac{M_H}{\mu_I},\alpha_s(\mu_I)\right)
&=&C_g(1,\alpha_s(M_H)) \nonumber \\
&\times& \exp\left(-\int^{M_H}_{\mu_I}
       \gamma_{1,g} \frac{d\mu}{\mu}\right) \ ,
\end{eqnarray}
where the exponent contains Sudakov double logarithms.

{\bf 3}. {\it Physics at scale $(1-z)M_H$}. At this scale, one
must consider soft-gluon radiations from the initial gluon
partons. In principle, one should formulate a soft-collinear
effective theory to calculate these contributions, as was done in
Ref. \cite{Man03}. However, this is unnecessary in practice and
the result can simply be obtained from a full QCD calculation at
the appropriate kinematic limit.

The real emission diagrams without any internal radiative
corrections contain both soft and collinear divergences. The
infrared contributions from diagrams with internal radiative
corrections (just the pole terms in dimensional regularization)
serve to cancel the soft divergences. The collinear divergences
can be factorized into a standard Feynman parton distributions.
The finite remainder is the coefficient function (or matching
coefficient in the sense that we match a product of effective
gluon currents onto a product of gluon light-cone distribution
operators) which is what we are interested in. To see that the
relevant physics happens around the scale $\mu_I$ is to take
Mellin transform of the coefficient function, which contains
logarithms of type $\alpha_s^k \ln^m M^2_H/\mu^2 \overline{N}^2$.
If $\mu$ is set as $\mu_I= M_H/\overline{N}$, the large logarithms
disappear.

Expanding the matching coefficient $ {\cal M}_N = \sum_i
a^i_s(\mu_I) {\cal M}_N^{(i)} $, the full pQCD calculation
\cite{Catani:2001ic,Anastasiou:2002yz} in the soft limit yields
\begin{eqnarray}
   {\cal M}_N^{(1)} &=& 2C_A \zeta_2 \nonumber \\
{\cal M}_N^{(2)}&=&
C_A^2\left[\frac{2428}{81}+\frac{67}{9}\zeta_2-\frac{22}{9}\zeta_3-10\zeta_2^2\right]\nonumber\\
&+&C_A
N_F\left[-\frac{328}{81}-\frac{10}{9}\zeta_2+\frac{4}{9}\zeta_3\right]
,
\end{eqnarray}
at scale $\mu_I=M_H/\overline{N}$. Note that the coupling constant
$\alpha_s$ is also evaluated at the intermediate scale $\mu_I$.

{\bf 4.} {\it Physics between scales $\mu_I$ and $\mu_F$, and at
$\mu_F$}. In the previous section, QCD factorization produces
gluon distributions at scale $\mu_I=M_H/{\overline N}$. We can
bring the distributions to an arbitrary scale $\mu_F$ using the
standard DGLAP (Dokshitzer-Gribov-Lipatov-Altarelli-Parisi)
evolution. This introduces an evolution factor, \begin{equation}
     \exp\left(2\int_{\mu_F}^{\mu_I} \frac{d\mu}{\mu}
       \gamma_{2,g}^N\right) \ ,
\end{equation}
where the twist-two anomalous dimension $\gamma_{2,g}^N$ has the
following large $N$ behavior:
\begin{equation}
     \gamma^N_{2,g} = -A_{g} \ln \overline{N}^2 +2
B_{2,g} \ ,
\end{equation}
where $A_{g}$ and $B_{2,g}$ are the same as those in Eqs. (8) and
(9), respectively.  The moment of gluon distributions introduces a
factor, \begin{equation}
         g(\mu_F,N)g(\mu_F,N) \ .
\end{equation}
To simplify the result, the factorization scale $\mu_F$ is
henceforth chosen to be $M_H$.

 {\bf 5.} {\it Resummation to all orders in
sub-leading logarithms}. Putting all factors together, the
$M_H^2/s$-moment of the cross section is ($s$ is the total
center-of-mass energy squared) \cite{Catani:2003zt}
\begin{equation}
   \sigma_N = \sigma_0\cdot G_N(M_H)\cdot
g(M_H,N)g(M_H,N) \ ,
\end{equation}
where $\sigma_0$ is a reference cross section and \begin{eqnarray}
   G_N(M_H) &=& |C(\alpha_s(M_H))|^2e^{I_1(M_H,\mu_I)}
\nonumber
   \\ && \times {\cal M}_N(\alpha_s(\mu_I))
   e^{I_2(\mu_I,M_H)}
\end{eqnarray}
is a pQCD factor, where $I_1 = 2\int_{M_H}^{\mu_I}
\frac{d\mu}{\mu}\tilde{\gamma}_{1,g}$ with
$\tilde{\gamma}_{1,g}=\gamma_{1,g}-2i\beta_{i-1}$, is the
anomalous dimension for $C=C_\phi\times C_g$, and $I_2 =
2\int^{\mu_I}_{M_H} \frac{d\mu}{\mu}\gamma_{2,g}$. To capture all
large $\ln\overline{N}$, we translate the dependence on the
intermediate scale $\alpha_s(\mu_I)$ to $\alpha_s(M_H)$, using
\begin{equation}
     {\cal M}_N(\alpha_s(\mu_I)) =
     {\cal M}_N(\alpha_s(M_H))~e^{I_3} \ ,
\end{equation}
where $I_3= -2\int^{M_H}_{\mu_I}
     \frac{d\mu}{\mu} \triangle B_1$, and $\triangle
B_1$ is defined as $ \triangle B_1=-\beta(a_s) d\ln {\cal
M}_N/d\ln \alpha_s$ .

The final form of resummation is
\begin{equation}
 G_N(M_H) = {\cal F}(\alpha_s(M_H)) e^{I(\lambda, \alpha_s(M_H))}
\end{equation} where ${\cal F} = |C(\alpha_s(M_H))|^2{\cal
M}(\alpha_s(M_H))$ depends only on $\alpha_s(M_H)$. In fact, it is
just the standard full-QCD coefficient function in the soft-gluon
approximation, without the large logarithms. $I=I_1+I_2+I_3$ is a
function of $\lambda=\beta_0\ln \overline N \alpha_s(M_H)$ and
$\alpha_s(M_H)$ with all leading and sub-leading large logarithms
resummed.

The above result can be related to the conventional expression if
one writes
\begin{equation}
     I = I_\Delta + \ln \Delta C \ ,
\end{equation}
where
\begin{eqnarray}
I_{\triangle} &=&
\int_0^1dz\frac{z^{N-1}-1}{1-z}\left[2\int_{M^2_H}^{(1-z)^2M^2_H}
\frac{d\mu^2}{\mu^2}A_g\left(\alpha_s(\mu^2)\right)\right.
   \nonumber \\ && \left.
+D_g(\alpha_s((1-z)^2M_H^2))\right] \ , \label{idelta}
\end{eqnarray}
and $\Delta C$ is just a function of $\alpha_s(M_H)$, serving to
cancel the non-logarithmic terms in $I_\Delta$. Using similar
methods as the ones in Ref. \cite{Catani:2003zt}, it is a matter
of some technical steps to get,
\begin{eqnarray}
D_g(\mu^2) &=&2( B_{1,g}+\Delta B_1+2B_{2,g}) \nonumber \\ && -
\partial_{\alpha_s}\Gamma_2(\partial_{\alpha_s})\left[4A_g(\alpha_s)-\partial_{\alpha_s}D_g(\alpha_s)\right]
\nonumber \\
  \Delta C &=&
\Gamma_2(\partial_{\alpha_s})\left[4A_g(\alpha_s)
   -\partial_{\alpha_s}D_g(\alpha_s)\right] \
, \label{master}
\end{eqnarray}
where $\Gamma_2(\epsilon)=
1/\epsilon^2[1-e^{-\gamma_E\epsilon}\Gamma(1-\epsilon)]
=-\zeta_2/2-\zeta_3\epsilon/3+...$ and
$\partial_{\alpha_s}=2\beta(\alpha_s)\alpha_s\partial/\partial\alpha_s$.
The above equations are our main result connecting the EFT
resummation to the coventional approach, valid to all orders in
leading and sub-leading logarithms. Similar results have been
obtained for deep-inelastic scattering and Drell-Yan processes
\cite{idilbi}.

{\bf 6.} {\it Results up to N$^3$LL and Conclusion}. The function
$I$ has a perturbative expansion in $\alpha_s(M_H)$,
\begin{eqnarray}
  I && = \ln N g_1(\lambda)
                + g_2(\lambda) + \alpha_s(M_H)
g_3(\lambda) \nonumber \\
    &&
    + \alpha_s^2(M_H) g_4(\lambda)
   + ... \ ,
\end{eqnarray}
where $g_i(\lambda)$ can be found from Ref.
\cite{Catani:2003zt,Moch:2005ba}. They are functions of $A_g$ and
$D_g$: $g_1(\lambda)$ sums over the leading logarithms, depending
on $A^{(1)}_g$, and $g_2(\lambda)$ sums over next-to-leading
logarithms, depending on $A^{(2)}_g$ and $D^{(1)}_g$, etc. The
$D_g^{(i)}$ coefficients can be solved iteratively from Eq.
(\ref{master}),
\begin{eqnarray}
D^{(1)}_g &=& 0 \nonumber \\
 D^{(2)}_g
&=&-2f_g^{(2)}+4\beta_0\zeta_2 A_g^{(1)}-2\beta_0{\cal M}_N^{(1)}
\nonumber \\ D^{(3)}_g&=&-2f_g^{(3)}+4\zeta_2\beta_1 A_g^{(1)}
+8\zeta_2\beta_0A_g^{(2)}+\frac{32}{3}\zeta_3\beta_0^2A_g^{(1)}
\nonumber \\ &&
 -2\beta_1 {\cal M}_N^{(1)}-2\beta_0\left[2{\cal M}_N^{(2)}-\left({\cal
M}_N^{(1)}\right)^2\right] \ , \end{eqnarray} and so on.

As an example to demonstrate the equivalence to the conventional
formalism, we calculate $D_g^{(3)}$ using the known result
$f_g^{(3)}$ \cite{Moch:2005tm}, ${\cal M}^{(1),(2)}$ (Eq. (13)),
and $A_g^{(1),(2)}$ \cite{Vogt:2004mw}. The answer is,
\begin{eqnarray} D_g^{(3)} &=&
C_A^3\left[-\frac{594058}{729}+\frac{98224}{81}\zeta_2+\frac{40144}{27}\zeta_3
\right.\nonumber \\ &&
\left.-\frac{2992}{15}\zeta_2^2-\frac{352}{3}\zeta_2\zeta_3-384\zeta_5\right]\nonumber\\
&+&C_A^2n_f\left[\frac{125252}{729}-\frac{29392}{81}\zeta_2-\frac{2480}{9}\zeta_3
+\frac{736}{15}\zeta_2^2\right]\nonumber\\
&+&C_AC_Fn_f\left[\frac{3422}{27}-32\zeta_2-
\frac{608}{9}\zeta_3-\frac{64}{5}\zeta_2^2\right]\nonumber\\
&+&C_An_f^2\left[-\frac{3712}{729}+\frac{640}{27}\zeta_2+\frac{320}{27}\zeta_3\right]\
,
\end{eqnarray}
which agrees completely with the recent calculations
\cite{Moch:2005ky,Laenen:2005uz}.

To conclude, we have presented an effective field theory method to
resum large threshold double logarithms in standard model Higgs
production. The approach is simple conceptually and uses the full
QCD calculation in the soft limit. The result is valid to all
orders in leading and sub-leading logarithms, and reproduces the
known answer to N$^3$LL order.

It has been shown that the inclusion of threshold resummation
effects helps to reduce the theoretical uncertainties in the
prediction of Higgs production rates at hadron colliders
\cite{Catani:2001ic}. It will be interesting to see if we can get
even better results when the newly calculated $D^{(3)}_g$ term
being included. We note that the other sources of theoretical
uncertainties, such as those stemming from heavy quark loop
approximation \cite{Dawson:1990zj,{Chetyrkin:1997iv}}, and those
from parton distribution parameterizations, need to be considered
in a detailed phenomenological studies of the Higgs production.

A. I. and X. J. acknowledge support of the U. S. Department of
Energy via grant DE-FG02-93ER-40762. X. J. and J. M. were
partially supported by National Science Foundation of China and
Ministry of Education of China. F.Y. thanks Iain Stewart and
Werner Vogelsang for useful discussions. F.Y. is grateful to
RIKEN, Brookhaven National Laboratory and the U.S. Department of
Energy (contract number DE-AC02-98CH10886) for providing the
facilities essential for the completion of his work.

\end{document}